# Magnetospheres of Black Hole Systems in Force-Free Plasma

Carlos Palenzuela[1,2], Travis Garrett[2,3], Luis Lehner[3,4,5] and Steven L. Liebling[6]

[1] *Canadian Institute for Theoretical Astrophysics, Toronto, Ontario M5S 3H8, Canada*
[2] *Dept. of Physics & Astronomy, Louisiana State University, Baton Rouge, LA 70803, USA*
[3] *Perimeter Institute for Theoretical Physics, Waterloo, Ontario N2L 2Y5, Canada*
[4] *Department of Physics, University of Guelph, Guelph, Ontario N1G 2W1, Canada,*
[5] *Canadian Institute For Advanced Research (CIFAR), Cosmology and Gravity Program, Canada*
[6] *Department of Physics, Long Island University, New York 11548, USA*
(Dated: July 7, 2010)

The interaction of black holes with ambient magnetic fields is important for a variety of highly energetic astrophysical phenomena. We study this interaction within the force-free approximation in which a tenuous plasma is assumed to have zero inertia. Blandford and Znajek (BZ) used this approach to demonstrate the conversion of some of the black hole's energy into electromagnetic Poynting flux in stationary and axisymmetric single black hole systems. We adopt this approach and extend it to examine asymetric and, most importantly, dynamical systems by implementing the fully nonlinear field equations of general relativity coupled to Maxwell's equations. For single black holes, we study in particular the dependence of the Poynting flux and show that, even for misalignments between the black hole spin and the direction of the asymptotic magnetic field, a Poynting flux is generated with a luminosity dependent on such misalignment. For binary black hole systems, we show both in the head-on and orbiting cases that the moving black holes generate a Poynting flux.

## I. INTRODUCTION

Enormously powerful events illuminate the universe that challenge our understanding of the cosmos. Indeed, intense energy outputs of order $10^{51}$ergs [1, 2] are routinely observed in supernovae and gamma ray bursts (GRBs) and in other puzzling systems, outputs as high as $10^{60}$ ergs have been inferred [3]. Despite important theoretical and observational advances, we still lack a complete understanding of systems such as active galactic nuclei (AGNs), GRBs, etc, and intense observational and theoretical efforts seek to unravel these fascinating phenomena. While the full details are still elusive, a natural ingredient in theoretical models is the inclusion of at least one spinning black hole which helps convert binding and rotational energy in a highly efficient process.

Our understanding of these systems benefits hugely from progress on the observation front. On the one hand, a proliferation of sensitive telescopes are providing a wealth of information gathered in the electromagnetic band. While on the other hand, the advent of sophisticated interferometric gravitational wave observatories and their continued upgrades promise to provide a new and complementary view of these events. Coincident detection of both electromagnetic and gravitational wave signals promises to revolutionize our understanding while at the same time leading to the refinement of theoretical models (e.g [4–7]).

The starting point for these theoretical models can be traced back to ideas laid out by Penrose [8] and Blandford and Znajek [9] to explain the extraction of energy from a rotating black hole. These seminal studies, along with subsequent work (see references in e.g. [10–12]), have provided a basic understanding of highly energetic emissions from single black hole systems interacting with their surroundings. While observations are consistent with these ideas, uncertainties remain due to insufficient knowledge of the physical parameters governing these systems. Gravitational waves from black hole systems should provide a much cleaner "picture" of the central engine, potentially revealing key properties. Binary systems with either two black holes or a black hole and a neutron star are copious producers of gravitational waves and should be observable with earth-based gravitational wave detectors out to $\simeq$ 1Gpc and with spaced-based detectors up to redshifts of $\simeq 5 - 10$ [13].

We therefore consider this basic picture of the extraction of energy from black holes in the more complex regime of binary black holes in the hope of revealing an emission mechanism complementary to gravitational waves. We concentrate on understanding possible Poynting flux emissions from both isolated black holes and binary black hole systems. Our goal is to elucidate the basic phenomenology arising from the interaction of single and binary black hole binaries interacting with plasma environments. This latter scenario may arise as a result of galaxy mergers [14, 15]. As the galaxies approach, their respective supermassive, central black holes form a binary in the merged galaxy. Through diverse interactions, the black hole binary tightens, hollowing out a region surrounded by a circumbinary disk. Eventually, its dynamics is governed by gravitational radiation reaction which ultimately leads to the merger of the black holes. The circumbinary disk will likely be magnetized, anchoring field lines, some of which will traverse the central region containing the binary. As a result of the ambient magnetic field, a low density plasma will surround the black holes [9] which will be affected by the binary black hole dynamics. This system, composed of two black holes, ambient magnetic field and tenuous plasma, can lead to a net Poynting flux from the system [16].

This work is organized as follows. We begin in Sec. II with a description of the equations and assumptions employed. Sec. III includes details of our numerical implementation, and Sec. IV describes our results for both single and binary black holes. In the former case we include a brief description of known solutions which we use to further test our implementation. We conclude with final comments in Sec. V. We discuss details of the Blandford-Znajek mechanism in the Appendix.

## II. FORMULATION AND NUMERICAL APPROACH

We solve the coupled Einstein-Maxwell system to model the black hole merger interacting with a force-free environment produced by an externally sourced magnetic field. The particular formulation for these systems has been discussed in detail in [17], so we only summarize briefly the main results, focusing on the treatment of charge and current following the force-free approximation.

### A. The Einstein equations

We adopt a Cauchy, or 3+1, formulation where the spacetime $(M, g_{ab})$ $(a, b = 0, 1, 2, 3)$ is foliated with spacelike hypersurfaces labeled by constant coordinate time $x^0 \equiv t = $ const. The intrinsic metric of these hypersurfaces is $\gamma_{ij} = g_{ij}$ $(i, j = 1, 2, 3)$. The normal vector to the hypersurfaces is $n_a \equiv -\nabla_a t / ||\nabla_a t||$, and coordinates defined on neighboring hypersurfaces can be related through the lapse function, $\alpha$, and shift vector, $\beta^i$. With these definitions, the spacetime line element can be expressed as

$$ds^2 = g_{ab} \, dx^a dx^b \\
= -\alpha^2 \, dt^2 + \gamma_{ij} \left(dx^i + \beta^i \, dt\right)\left(dx^j + \beta^j \, dt\right). \quad (1)$$

The normal vector/covector are given explicitly by

$$n^a = \frac{1}{\alpha}(1, -\beta^i) \; , \; n_a = (-\alpha, 0) \; . \quad (2)$$

Indices on spacetime quantities are raised and lowered with the 4-metric, $g_{ab}$, and its inverse, while the 3-metric, $\gamma_{ij}$, and its inverse are used to raise and lower indices on spatial quantities. The following simple expressions relate the 3+1 basic variables $\{\gamma_{ij}, \alpha, \beta^i\}$ with the four-dimensional metric $\{g_{ab}\}$ by

$$\gamma_{ij} = g_{ij} \; , \; \alpha = \sqrt{-1/g^{00}} \; , \; \beta^i = \gamma^{ij} g_{0j} \; . \quad (3)$$

We adopt Einstein's equations written in the Generalized Harmonic (GH) formulation to evolve the full space-time metric $g_{ab}$. We adopt a fully first order formulation of the GH equations together with constraint damping as described in [17–20]. The 3 + 1 variables $\{\gamma_{ij}, \alpha, \beta^i\}$ are employed to express Maxwell equations in a more familiar form.

### B. Maxwell equations

To implement the Maxwell equations, we adopt the formulation described in [17, 21]. The equations of motion for the electric and magnetic fields are given by

$$(\partial_t - \mathcal{L}_\beta) E^i - \epsilon^{ijk} \nabla_j(\alpha B_k) + \alpha \gamma^{ij} \nabla_j \Psi = \\
\alpha \, \text{tr}K \, E^i - 4\pi \alpha J^i \,, \quad (4)$$

$$(\partial_t - \mathcal{L}_\beta) B^i + \epsilon^{ijk} \nabla_j(\alpha E_k) + \alpha \gamma^{ij} \nabla_j \phi = \\
\alpha \, \text{tr}K \, B^i \,, \quad (5)$$

$$(\partial_t - \mathcal{L}_\beta) \Psi + \alpha \nabla_i E^i = 4\pi \alpha \, q - \alpha \sigma_2 \, \Psi \,, \quad (6)$$

$$(\partial_t - \mathcal{L}_\beta) \phi + \alpha \nabla_i B^i = -\alpha \sigma_2 \, \phi \,. \quad (7)$$

with trK the trace of the extrinsic curvature, $q$ the charge density and $J^i$ the current density. The fields $\Psi$ and $\phi$ are introduced to dynamically enforce the constraints via an exponential damping in a characteristic timescale $1/\sigma_2$.

In previous works, we have employed this formulation to study electrovacuum scenarios [17, 22, 23]. We are here interested in the more realistic case that considers plasma around the black holes. To this end, we recall that in the magnetospheres of the neutron stars or black holes the density of the plasma is so low that even moderate magnetic fields stresses will dominate over the pressure gradients. In turn, this means that the stress-energy tensor is dominated mainly by the electromagnetic component

$$T_{\mu\nu} = T_{\mu\nu}^{\text{fluid}} + T_{\mu\nu}^{\text{em}} \approx T_{\mu\nu}^{\text{em}} ; \quad (8)$$

the stress-energy conservation law implies that the Lorentz force is negligible. This is known as the *force-free approximation* [9, 24, 25], which can be written in terms of Eulerian observers as

$$E^k J_k = 0 \quad , \quad q E^i + \epsilon^{ijk} J_j B_k = 0 \; . \quad (9)$$

By considering the scalar and the vectorial products with the magnetic field $B^i$ in Eq. (9), one obtains

$$E^i B_i = 0 \; , \quad (10)$$

$$J^i = q \frac{\epsilon^{ijk} E_j B_k}{B^2} + J_B \frac{B^i}{B^2} \; , \quad (11)$$

where $J_B \equiv J^k B_k$ is the component of the current parallel to the magnetic field. The first relation implies that the electric and magnetic fields must be perpendicular while the second defines the current up to the parallel component $J_B$. By using Maxwell equations, one can compute $\partial_t(E^i B_i)$, which has to vanish due to the constraint in Eq. (10). This condition imposes a relation for $J_B$, which can be substituted into Eq. (11) in order to complete the specification of the current (e.g. [25, 26]).

An alternative approach to determine the parallel component is through the introduction of a suitable Ohm's law of the type

$$J_B = \sigma_B E^k B_k, \quad (12)$$

where $\sigma_B$ is the anisotropic conductivity along the magnetic field lines. Once the current is complete, we can use the Maxwell equations to compute again the time derivative of Eq. (10), obtaining

$$\partial_t(E^i B_i) = ... - \alpha\, \sigma_B (E^i B_i) \quad . \quad (13)$$

which enforces the constraint of Eq. (10) in a timescale given by $1/\sigma_B$. In the case of force-free plasmas, one has the limit $\sigma_B \to \infty$.

The above describes the basic aspects of the force-free equations which we use to model our systems of interest. In this approach, we are free to eliminate the charge density $q$ from our set of variables by substituting for it with the electromagnetic constraint $q = \nabla_i E^i$. Since we explicitly make use of the constraint, there can be no violation and so the divergence cleaning scalar field corresponding to the constraint on the electric field reduces trivially to $\Psi = 0$ (for further details on the implementation of divergence cleaning techniques see [17]). Furthermore, since in the force-free limit the inertia of the fluid is neglected, the fluid equations need not be evolved at all. It is useful to note that the characteristic speeds of the Maxwell equations in the force-free limit are given by two Alfvén waves and two magnetosonic waves, moving at the speed of light.

One last delicate point is that the force-free approximation may break down during the evolution in some regions. For instance, in some regions current sheets develop such that

$$B^2 - E^2 > 0 \quad . \quad (14)$$

is not satisfied everywhere. Ideally, in such regions one expects anomalous isotropic resistivity to appear, restoring the dominance of the magnetic field. However, implementing some form of Ohm's law that accounts for this resistivity will generally lead to stiff terms in the evolution equation of the electric field. As a result, a severe constraint on the size of time step is introduced as $\Delta t \leq 1/\sigma_B$. This condition is computationally prohibitive in realistic scenarios and generally requires specialized algorithms to solve it. One possible approach is the use of Implicit-Explicit (IMEX) Runge-Kutta methods [27], although its implementation in general relativistic settings is far from mature.

We here adopt a different, simpler, option which consists of implementing these two resistivities (ie, the infinite anisotropic one and the anomalous isotropic one) by following a prescription given in [21, 28]. We evolve the Maxwell equations with the component of current perpendicular to the magnetic field given in Eqs. (11), but, after each timestep, we modify the resulting electric field to account for the role the resistivity would play. In particular, we modify the electric field so that it satisfies both $E^i B_i = 0$ and $B^2 - E^2 > 0$. These conditions are enforced by suitably projecting $E^i$ as,

$$E^i \to E^k \left(\delta^i_k - B_k \frac{B^i}{|B|^2}\right) \quad (15)$$

$$E^i \to E^i \left[\left(1 - \Theta(\chi)\right) + \frac{|B|}{|E|}\Theta(\chi)\right] \quad (16)$$

where $\chi = E^2 - B^2$ and function $\Theta(\chi) = 1$ if $\chi > 0$ and 0 otherwise. After each iteration of the Runge-Kutta time step, Eq. (15) is applied to remove any component parallel to the magnetic field and then Eq. (16) is applied to limit the magnitude of the electric field to that of the magnetic field.

## III. IMPLEMENTATION

### A. Numerical Implementation

We adopt finite difference techniques on a regular Cartesian grid to solve the overall system numerically. To ensure sufficient resolution in an efficient manner we employ adaptive mesh refinement (AMR) via the HAD computational infrastructure that provides distributed, Berger-Oliger style AMR [29, 30] with full sub-cycling in time, together with an improved treatment of artificial boundaries [31]. The refinement regions are determined using truncation error estimation provided by a shadow hierarchy [32] which adapts dynamically to ensure the estimated error is bounded by a pre-specified tolerance. A fourth order accurate spatial discretization satisfying a summation by parts rule together with a third order accurate in time Runge-Kutta integration scheme are used to help ensure stability of the numerical implementation [33]. We adopt a Courant parameter of $\lambda = 0.2$ so that $\Delta t_l = 0.2 \Delta x_l$ on each refinement level $l$. On each level, one therefore ensures that the Courant-Friedrichs-Levy (CFL) condition dictated by the principal part of the equations is satisfied.

To extract physical information, we monitor the Newman-Penrose radiative scalars; in particular, the electromagnetic ($\Phi_2$) and gravitational ($\Psi_4$) radiative scalars [?]. These scalars are computed by contracting the Maxwell and the Weyl tensors respectively, with a suitably defined null tetrad

$$\Phi_2 = F_{ab} n^a \bar{m}^b \quad , \quad \Psi_4 = C_{abcd} n^a \bar{m}^b n^c \bar{m}^d \quad ; \quad (17)$$

and they account for the energy carried off by outgoing waves at infinity. The total energy flux (luminosity) in both electromagnetic and gravitational waves are

$$L_{EM} = \frac{dE^{EM}}{dt} = \int F_{EM} d\Omega$$

$$= \lim_{r\to\infty} \int \frac{r^2}{2\pi} |\phi_2|^2 d\Omega \ , \tag{18}$$

$$L_{GW} = \frac{dE^{GW}}{dt} = \int F_{GW} d\Omega$$

$$= \lim_{r\to\infty} \int \frac{r^2}{16\pi} \left| \int_\infty^t \Psi_4 dt' \right|^2 d\Omega \ . \tag{19}$$

### B. The Blandford-Znajek model

Seeking to understand the nature of the central engine powering AGNs, Blandford and Znajek (BZ) studied the extraction of rotational energy from a spinning BH by means of an electromagnetic field [9]. Their model assumes a spinning black hole immersed in an magnetic field produced by a magnetized accretion disk. The rotation of the BH within the magnetic field induces a charge separation between the poles and the equator of the BH horizon (which can be understood easily by the membrane paradigm [34]), producing a potential difference in the immediate vicinity of the black hole. A single electron (or positron) accelerated by this potential difference will reach a high enough energy to radiate gamma-ray photons, which in turn may decay into an electron-positron pair. This pair production process can be repeated, leading to a cascade. The time-averaged structure of this magnetosphere is reasonably well described by the force-free approximation.

Another important diagnostic quantity measures the behavior of the electromagnetic field. Let us write down the Maxwell tensor in terms of the vector potential $A_a$, namely

$$F_{ab} = \partial_a A_b - \partial_b A_a \ , \tag{20}$$

and assume that the spacetime is axisymmetric and stationary, that is, $\partial_\phi F = \partial_t F = 0$ for any field $F$. The force-free condition $E^i B_i = 0$ can be written as $^*F^{ab} F_{ab} = 0$, or in terms of the vector potential and standard spherical coordinates

$$A_{\phi,\theta} A_{t,r} - A_{t,\phi} A_{\phi,r} = 0 \ . \tag{21}$$

We can define a function $\Omega_F(r,\theta)$ such that

$$\Omega_F \equiv -\frac{A_{t,r}}{A_{\phi,r}} = -\frac{A_{t,\theta}}{A_{\phi,\theta}} \ , \tag{22}$$

which can be interpreted as the rotation frequency of the electromagnetic field. Because the poloidal field surfaces can be defined by $A_\phi = $ constant (i.e., it is a stream function for the magnetic field), $\Omega_F$ and the electrostatic potential $A_t$ are therefore constant along magnetic field lines. Notice that $\Omega_F$ can also be written in terms of the Maxwell tensor, in particular

$$\Omega_F = \frac{F_{tr}}{F_{r\phi}} = \frac{F_{t\theta}}{F_{\theta\phi}} \ . \tag{23}$$

The quantity $\Omega_F$ thus represents a useful quantity which we monitor in our simulations. We emphasize however that $\Omega_F$ is defined strictly in terms of a stationary, axisymmetric spacetime, and so it need not be useful for the dynamical or asymmetric cases that we study in what follows.

Traditionally, two lines of thought have been adopted when describing the ability to extract energy from a black interacting with a magnetic field in the force-free approximation. One of them exploits the assumption of stationarity to calculate the amount of energy extracted from the system while the other appeals to the membrane paradigm to interpret the system as a loaded circuit which dissipates energy. (For reference we include a brief overview of both approaches in the appendix). Notice that both options suffer from caveats in their interpretation; however their basic picture and message is the same. Namely that a net flux of electromagnetic energy is produced the magnitude of which scales as $\propto (B^r)^2 \Omega_F (\Omega_H - \Omega_F)$ with $\Omega_H \equiv a/(2Mr_H)$ the frame dragging orbital frequency at the horizon, $B^r$ the normal component of the magnetic field at the horizon, and $r_H$ is the horizon radius of the BH. Without a known solution for the cases of interest, both $B^r$ and $\Omega_F$ need to be obtained from numerical solutions. In cases studied [9], it has been found that $\Omega_F \simeq \Omega_F/2$ for the monopole case, and also at the poles in the case of a black hole immersed in an otherwise constant magnetic field aligned with the black hole spin (discussed more in Sec. IV A 1).

### C. Initial Data

We consider both single and binary black hole simulations, immersed initially in a constant magnetic field such as one produced by a distant disk surrounding the black hole. Because the electromagnetic field is affected by the curved spacetime, it will be dynamically distorted from its initial configuration, but will eventually reach a quasi-stationary configuration.

In addition to its direction, we must choose an initial magnitude $B_0$ for the magnetic field, and we choose an astrophysically relevant value. To this end, we first express the magnitude in geometrized units

$$B[1/M] = 1.2 \times 10^{-20} \left( \frac{M}{M_\odot} \right) B[G] \ . \tag{24}$$

We adopt (except in the monopole case described in Sec. IV A 1) a field strength of $B_0 = 10^4 (M/10^8 M_\odot)$ G, which is consistent with possible values inferred in relevant astrophysical systems [35, 36] and which is below the Eddington magnetic field strength $B \simeq 6 \times 10^4 (M/10^8 M_\odot)^{-1/2} G$ [37]. Such realistic magnitudes for the magnetic field dictate that the energy associated with the electromagnetic field remains several orders of magnitude smaller than that of the gravitational field and so they have a negligible influence on the dynamics of the black holes. For all binary cases considered here, the orbital plane of the evolution is assumed to be aligned with that of the distant circumbinary disk. Because the magnetic field is anchored in the disk, its associated magnetic dipole is aligned with the orbital angular momentum. The electric field is always chosen to be initially zero.

For the single black hole cases, we adopt the Kerr spacetime written in horizon penetrating (i.e., Kerr-Schild) coordinates. A superposition of two isolated black holes in these coordinates is used for the head-on binary black hole case. The individual mass of each black hole is $M_s = 2$, and they are initially at rest and separated by a distance of $16\,M$, where $M = 2\,M_s$ the total mass.

For orbiting binary black holes, we adopt initial data corresponding to quasi-equilibrium, equal-mass, non-spinning black holes constructed by the publicly available LORENE code [38]. The black holes are initially separated by a distance of $\approx 6M$, lying beyond the approximate inner most stable circular orbit (ISCO) [39]. With this separation, the merger takes place after about one orbit and we can compare with the results obtained in the electrovacuum case [17, 22].

## IV. NUMERICAL RESULTS

The analysis of the single black hole case serves not only as a test of our numerical implementation, but also as a more basic system with which to understand the orbiting case. In particular, the features of the initial transient, in which the EM field adapts to the geometry of the black hole spacetime, gives rise to an electric field and deforms the initially vertical magnetic field (see also [40, 41]). Notice however that the electric and magnetic fields are not coordinate invariant quantities [?]. Therefore, in general they will have different forms in different coordinate systems and, as always, one should proceed with care when examining non-invariant quantities.

To explore the effects of the merger dynamics on the electromagnetic field, we compare single spinning black hole with cases of equal-mass merging black holes. For single black hole cases, we examine the behavior of the system with respect to variations on the spin parameter as well as different inclinations between spin and asymptotic magnetic field directions.

### A. Single black holes

We consider two cases involving single black holes as tests of our implementation. We note our use of Cartesian coordinates which are not adapted to the symmetries of these two test cases, and so our angular resolution is generally not as refined as in other work with 2D axisymmetric codes [42]. However, comparison of our results with such work gives us further confidence on our implementation. In these simulations of single black holes the geometry is kept fixed.

Our numerical domain consists of a right parallelepiped region defined by the intervals $[-40\,M, 40\,M]^2 \times [-100\,M, 100\,M]$. This structure allows us to refine further along the $z$-direction to better resolve the resulting collimated Poynting flux. We employ a fixed mesh refinement (FMR) configuration with 5 levels of refinement, each one covering half of the domain of the parent, coarser level. The coarsest resolution employed is $\Delta x^i = 2M$ while the finest one is $\Delta x^i = 0.125M$. The damping parameter is set to be $\sigma_2 = 1M$.

#### 1. The monopole solution

There is an exact, flatspace solution of Michel [43] for a non-rotating black hole with purely radial magnetic field [9, 21]

$$B^r = B_0 \sin\theta/\sqrt{\gamma} = \alpha B_0/r^2 \,. \qquad (25)$$

For scenarios with $a \ll 1$, the poloidal magnetic field is not expected to differ much from this exact solution because the differences scale with $O(a^2)$. Blandford and Znajek found a perturbative solution for slow rotation which demonstrates rotation of the magnetic field lines [9]. Matching this solution for large radius with Michel's monopole solution in flat spacetime leads to $\Omega_F = \Omega_H/2$; the magnetic field lines rotate with constant angular velocity everywhere at half of the rotation frequency of the BH.

We adopt $B_0 = 0.01$ in geometrized units and $a = 0.1$, since for this low spin the numerical solution results are close to the perturbative one. The initial electric field is set initially to zero, although it evolves gradually to the Blandford-Znajek monopole solution during the evolution. Fig. 1 displays the magnetic field lines on the $z = 0$ plane and the angular distribution of $\Omega_F$ in the azimuthal and axial directions at $t = 100M$. The results are in good agreement with the perturbative solution and compare well with Komissarov (see Fig. 2 of [21]).

#### 2. The force-free "Wald solution"

In the absence of charges and currents (i.e., the electrovacuum case), an exact solution was constructed by Wald [44] for a black hole immersed in an external magnetic field aligned with the spin. Although an analogous exact solution for the force-free approximation is unknown, several numerical studies have made use of Wald's solution as initial data for a force-free evolution [21, 45]. These studies found that all magnetic field lines penetrating the ergosphere rotate with a frequency similar to the paraboloidal case of Blandford-Znajek [9].

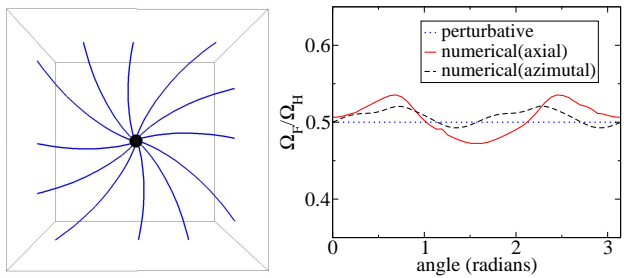

FIG. 1: *Monopole test:* **Left:** Magnetic field lines in the $z = 0$ plane at $t = 100M$ for a single spinning black hole with $a = 0.1$. **Right:** The rotation frequency $\Omega_F$ of the magnetic field along the axial and azimuthal angles about the origin. The known perturbative solution is also shown for comparison purposes.

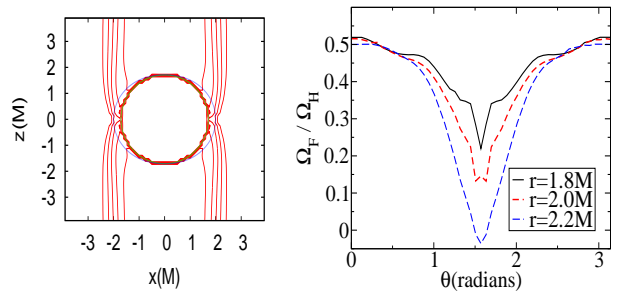

FIG. 2: *"Wald solution":* Late time ($t = 100M$) behavior for a single spinning black hole with $a = 0.7$. **Left:** Contour lines of the rotation frequency $\Omega_F/\Omega_H$ between 0.01 and 0.9, together with the apparent horizon (green) and the ergosphere (blue). **Right:** The rotation frequency interpolated onto the half-circle in a meridional plane.

We revisit this setup and consider different spin values and alignment. In the latter case we adopt a spin parameter given by $a = 0.7M$ which is close to the spin expected for a merged black hole from an equal-mass, non-spinning binary system. We adopt this value for comparison with the binary black hole scenario presented in the next section.

The initial magnetic field is poloidal resulting from a circular current loop, the radius of which is assumed to be larger than the region of interest [46]. We assume the disk lies at $10^3 M$, and for these distances the magnetic field is essentially constant and vertical within our computational domain. We therefore simply set $B^i = B_0 \hat{z}$ (with $B_0 = 10^4$ G) with an initially vanishing electric field.

The evolution shows an initial transient during which the magnetic field twists around the spinning black hole and an electric field is induced. After $t \simeq 80M$ the solution evolves towards a quasi-stationary state. As displayed in Fig. 2, all magnetic field lines crossing the ergosphere acquire a rotation velocity consistent with previous studies [21, 45].

We compare the resulting electrovacuum and force-free solutions at late time in Fig. 3. From the figure, it is quite apparent that the force-free solution results in significantly more deformation of the magnetic field lines because of the resulting currents near the horizon. Similarly, the electric field of the force-free case reveals a "flow" structure on the black hole which can support currents along it. In contrast, the electrovacuum case has an induced separation of charge but no current. The presence of charge and current in the force-free model is the critical difference which allows for the large Poynting flux and energy extraction from the black hole.

*3. The Dependence of Luminosity on Spin and Orientation*

The single black hole, outside the scope of any test, represents a physically interesting scenario in and of it-self, and it also describes the late stage of dynamical processes resulting from the merger of black holes. In particular, we consider the dependence of the radiated power computed in Eq. (18) on the spin of the black hole. Fig. 4 (top panel) shows the results of varying the spin parameter $a$ along with a fit obtained by the following argument. Following the analysis of [47] for the monopole solution, we have

$$\Omega_F \simeq \frac{\Omega_H}{2} \quad , \quad B^r \simeq \frac{\Phi_T}{r^2} \, , \qquad (26)$$

where $\Phi_T$ is the total magnetic flux threading the black hole. Substituting these into Eq. (30), we obtain that the EM luminosity behaves as $L_{EM}(a) \simeq k\,\Phi_T^2\,\Omega_H^2$ for some constant $k$. As shown in Fig. 4, the dependence on $\Omega_H^2$ fits the numerical results quite well. Higher order expansions developed in [47] show that the next leading order goes like $\Omega_H^4$.

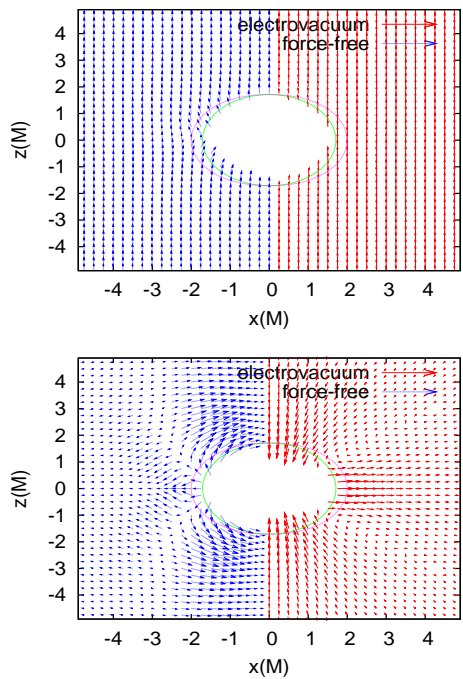

FIG. 3: Comparison of electrovacuum and force-free solutions: Snapshot at $t = 100M$ of a single spinning black hole with $a = 0.7$ on the $y = 0$ plane with the apparent horizon (green) and the ergosphere (magenta). For $x < 0$, we show the force-free solution while for $x > 0$ the electrovacuum solution is shown. **Top:** Magnetic field lines. **Bottom:** Electric field lines.

Of perhaps more interest, we consider the dynamics when the spin of the black hole is not aligned with the asymptotic direction of the magnetic field. We define the angle between these two directions as $\theta_0$ such that the aligned case is $\theta_0 = 0$. Varying $\theta_0$ up to $\pi/2$ radians breaks axial symmetry which we can study since our implementation does not assume any symmetry [?]. Fig. 4 (bottom panel) illustrates the observed luminosity for two cases with spin parameter fixed at $a = 0.7$ and $a = 0.1$ but misaligning it with respect to the asymptotic direction of the magnetic field. The emitted power decreases gradually for larger angles but even for the extreme case where the black hole spin is orthogonal to the asymptotic magnetic field, the power output is decreased to only about half that of the aligned case. Thus regardless of the spin orientation with respect to the magnetic field direction, a significant Poynting flux arises.

The observed dependence with inclination can be understood, in the slowly spinning limit, by combining the understanding of the BZ mechanism (within the membrane paradigm point of view) together with results of [48]. The work of [48] considers a slowly spinning and misaligned black hole within which currents are induced as it spins within the ambient magnetic field. Integrating the induced electric field, one obtains the electromotive force $\Delta V$ which is consistent with the black hole serving as a battery (also discussed in the Appendix). The resulting energy released at the load far outside the black hole then behaves as $\propto \left(1 + \cos(\theta_o)^2\right)$. Fig. 4 (bottom panel) illustrates the values obtained for spins $a = 0.1, 0.7$ for a range of angles $\theta_o$. We also include for comparison the curve given by $A_0(1 + \cos(\theta_o)^2)$ where we obtain the value of the fitting constant $A_0$ by matching to the value obtained for $a = 0.1$ at $\theta_o = 0$. For $a = 0.1$ good agreement is observed, especially for $\theta_o < 45^o$, suggesting the dominant behavior is indeed captured by the argument above. Because truncation errors in the calculated values are larger for higher angles, the discrepancy observed should not be taken too seriously before further refined results are available. Interestingly, the data for $a = 0.7$ demonstrates a similar trend, but a small bump arises around $\theta_o \approx 15^o$. Further work is required to assess whether this effect is real or driven by numerical errors.

It is instructive to examine the induced currents in both the aligned and orthogonal cases. Fig. 5 shows the currents within the $\hat{x}$-$\hat{z}$ plane, demonstrating that in both cases the current flows along the $z$-axis, following the background $B_0 \hat{z}$ field. For the aligned case, the current flows in a region near the $z$-axis down to the black hole, and then back up within a cylindrical shell at larger radius. The currents in the orthogonal case resemble those in the aligned case after antisymmetrization along the central axis.

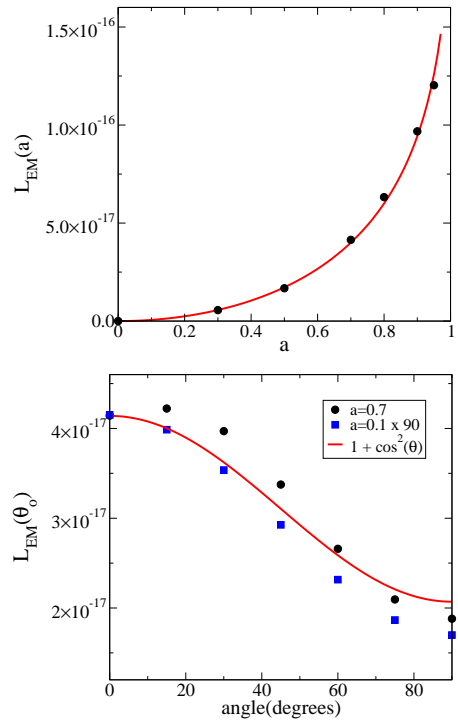

FIG. 4: Dependence of luminosity on spin and orientation for a black hole with mass $M = 10^8\,M_\odot$ and initial magnetic field of $B_0 = 10^4$ G. **Top:** The luminosity is shown as a function of the black hole spin $a$ for the aligned case $\theta_0 = 0$. Also shown (red, solid line) is a fit to $L_{EM} \propto \Omega_H^2$. **Bottom:** Luminosity as a function of the angle $\theta_0$ between the asymptotic magnetic field and the spin angular momentum of the black hole for the cases $a = 0.7$ and $a = 0.1$. Also shown (red, solid line) is the curve $A_o \left[1 + \cos(\theta_o)^2\right]$ where $A_o$ is a constant obtained by matching just the point at $\theta_o = 0$.

Fig. 6 shows cross sections of the magnetic field structures for the two alignments. We have switched to the $\hat{x}$-$\hat{y}$ plane at a distance $z = 4M$ above the black hole, so that the current $J^z$ now flows perpendicular to the diagrams. The aligned case ($\theta_o = 0$) shows the standard toroidal $B$ field with clockwise rotation. For the anti-aligned case at $\theta_o = \pi$ one would find the mirror image, with counterclockwise rotation. For the intermediate angles $0 < \theta_o < \pi$ however, this simple, single toroidal structure is not possible. In particular, directly flipping the helicity for a single toroidal $B$ field would result in irrotational field at $\theta_o = \pi/2$, and thus zero current and energy flux for this alignment. Instead we find that the system responds by generating two counter-rotating toroidal $B$ fields at $\theta_o = \pi/2$, offset by about the diameter of the black hole. It is this structure which leads to the antisymmetric currents seen in Fig. 5, and allows for a smooth transition from $\theta_o = 0$ to $\pi$ while always keeping $\nabla \times \vec{B}$ nonzero.

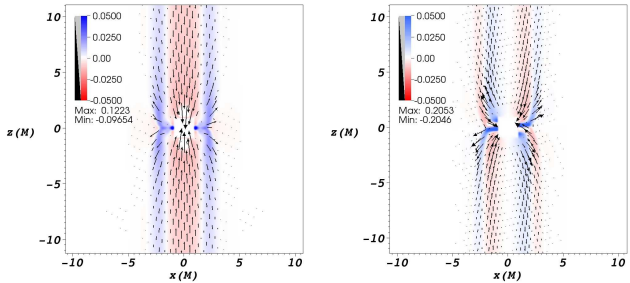

FIG. 5: Charge density and currents at $t = 100M$ for a single spinning black hole with $a = 0.7$. **Left:** An aligned black hole ($\theta_o = 0$). **Right:** A misaligned black hole with $\theta_o = \pi/2$.

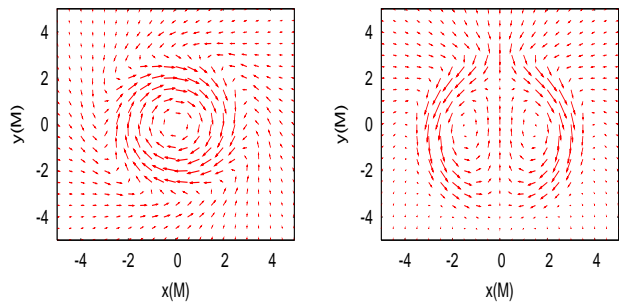

FIG. 6: Cross sections of the magnetic field structure at $z = 4M$ for the aligned and 90 degree cases. **Left:** A clockwise toroidal $B$ field is generated in the aligned case. **Right:** We find two offset and counter-rotating toroidal fields for $\theta_o = \pi/2$.

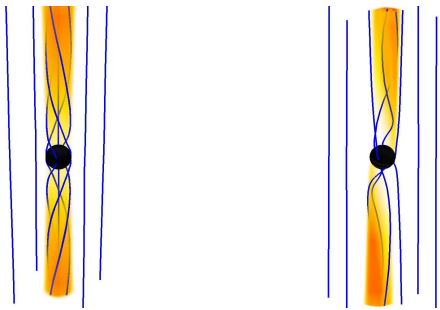

FIG. 7: Density of energy flux $F_{EM}$ at $t = 100M$ for a single spinning black hole with $a = 0.99$, together with the magnetic field lines. **Left:** An aligned black hole ($\theta_o = 0$). **Right:** A misaligned black hole with $\theta_o = \pi/4$.

Finally, as an illustrative example of the collimation characteristic of the energy output by the system, we display the energy flux density in Fig. 7 together with the magnetic field lines for two cases, $\theta_o = 0$ and $\theta_o = \pi/4$. The spin is set to a large value, $a = 0.99$, in order to demonstrate the twisting of the magnetic field lines. Clearly, the Poynting flux is directed along the field lines and is concentrated on the magnetic field lines which pass through the ergosphere.

### B. Binary Black Holes

In the cases with single black holes studied above, the energy was extracted along the field lines that cross the ergosphere. The monopole solution, therefore, radiates in all directions since all field lines cross the ergosphere. In contrast, the force-free "Wald solution" demonstrates a collimated energy flux along the direction of the magnetic field. This collimation is induced even when the spin of the black hole is orthogonal to the asymptotic magnetic field. These results lead one to consider non-spinning black holes that are nevertheless moving with respect to the asymptotic magnetic field.

We therefore now consider binary black hole cases, beginning with the simpler case of a head-on collision and following with an orbiting case. In all cases for simplicity we adopt equal mass binaries. The initial magnetic field is chosen once again as a poloidal configuration produced by a circular loop with large radius, so that $\vec{B} = B_0 \hat{z}$. The electric field is initially zero throughout the computational domain and the magnetic field strength adopted is $B_0 = 10^4$ G.

#### 1. Head-on collision

The domain is a cube given by $[-66\,M, 66\,M]^3$ and we employ an AMR configuration with 4 levels of refinement in which the coarsest resolution is $\Delta x^i = 1.5M$ while the finest one is $\Delta x^i = 0.09375M$. We adopt the same gauge parameters as in [17, 22]. The black holes begin stationary, accelerate towards each other and eventually merge. As time progresses the black holes increase in speed and eventually merge.

As illustrated in Fig. 8, the field lines are pulled by the motion of black holes. Because the field lines are fixed asymptotically by the distant circumbinary disk, the field lines cannot simply move with the hole. Instead, the local distortion of the field lines represent a complicated interplay among the black hole, the electromagnetic field, and the current of the plasma. The net result of this interaction is that some energy of the black hole is propagated by means of Alfvén waves.

Evident in Fig. 8 is a region of flux associated with each black hole. This flux is not associated with a rotation of the magnetic field lines, and so it is qualitatively different from what we find with single, spinning black holes. It is also important to note that we are venturing away from the scenario described by [9], since we have two non-spinning black holes with no angular momentum. There is therefore no ergoregion. Instead, the motion of the black holes with respect to the preferred frame of the magnetic field is serving to convert gravitational to electromagnetic energy. However, this process can be understood in terms similar to those used by BZ to describe single black holes. In particular, the motion of the black holes through the background magnetic fields induces an electromotive force –as in any circuit moving through a magnetic field. As the black holes merge, the final one is stationary with respect to the asymptotic field lines and does not spin, therefore its Poynting flux decreases to zero.

Finally, the structure and sign of the charge densities can be observed in Fig. 9. The membrane paradigm predicts a separation of charges on the apparent horizon analogous to the Hall effect. Notice a small current sheet develops behind the black holes.

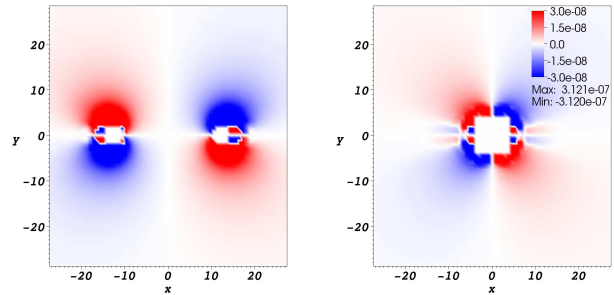

FIG. 9: *Head-on collision of two black holes:* The charge density before and at the merger on the equatorial plane.

### 2. Orbiting black holes

We now turn our attention to an orbiting binary black hole case. We adopt a cubical domain given by $[-106\,M, 106\,M]^3$ and employ an AMR configuration with 6 levels of refinement. The coarsest grid has just 47 points along each direction, so that the coarsest resolution is $\Delta x^i = 4.6M$ while the finest one is $\Delta x^i = 0.072M$. We adopt the same gauge parameters as in [17, 22].

The black holes rotate around each other for about an orbit before merging. Snapshots of this process are shown in Fig. 10. Even though the black holes have no initial spin, we observe a collimated region of Poynting flux associated with each of the black holes while in orbit. After merger, as the remnant black hole settles into a stationary, Kerr configuration, the Poynting flux likewise settles into what appears to be the standard Blandford-Znajek scenario. This result indicates that the B-Z scenario is a stable, attracting solution with no assumed symmetries and with dynamical gravity.

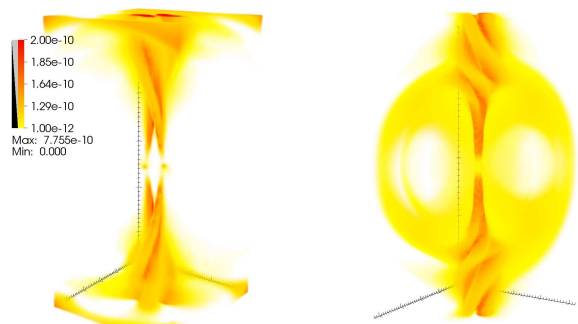

FIG. 10: *Orbiting binary black holes:* The electromagnetic energy flux at different stages of the evolution: early when the black holes are separated; shortly after they merge. Notice that the final state is similar to the single spinning black hole studied previously.

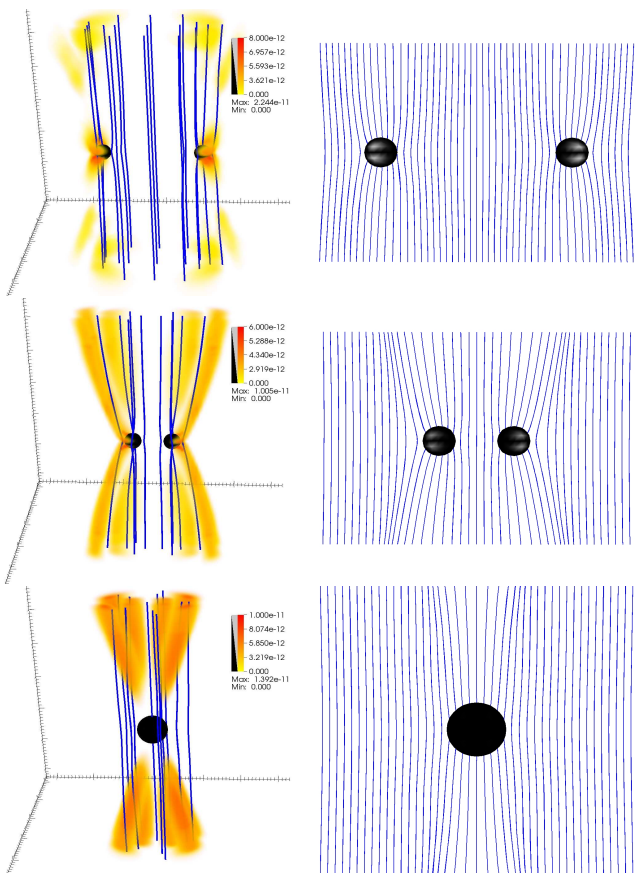

FIG. 8: *Head-on collision of two black holes:* Snapshots at times $t = 20M, 40M$ and $60M$. **Left:** The electromagnetic energy flux (shaded) and representative magnetic field lines (solid blue lines). **Right:** The magnetic field lines in detail on the $x = 0$ plane, showing only field lines close to the black holes are being perturbed, and that, after the BH passes, they recover their original shape due to the magnetic tension.

The electromagnetic and the gravitational luminosities throughout the merger are shown in Fig. 11. These luminosities are obtained with Eqs. (18) and (19) by inte-grating over a large sphere enclosing the system (radius $R_\Sigma = 20M$). As can be discerned in the figure, the luminosities increase dramatically at merger. It is also interesting that the gravitational wave luminosity is so much larger than the electromagnetic emission that it has to be scaled down to appear on the graph.

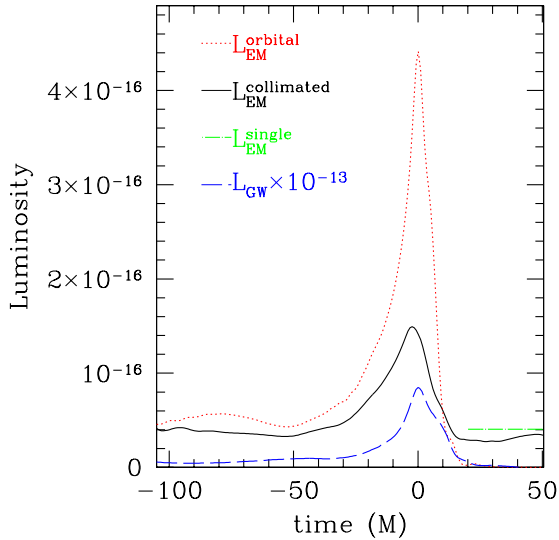

FIG. 11: Radiated power in the gravitational and electromagnetic band corresponding to the binary black hole case. Three electromagnetic luminosities; the collimated part (red), the corresponding to the $m = 2$ and so induced by the orbital motion (black) and the expected for the final spinning black hole with $a = 0.7$ (green). The isotropic part can be estimated by the subtracting the collimated to the orbital one. The gravitational power (blue) has been rescaled by a factor $10^{-13}$ to make it appear clearly in the plot.

We are interested in estimating the amount of luminosity, $L_{\rm EM}^{\rm isotropic}$, that *is not* collimated because such emission could potentially be visible when the jet is not pointing at us. This estimation, however, is delicate because the magnetic field is not localized which obscures the luminosity computation.

So we proceed by trying to isolate various contributions to the electromagnetic luminosity. First, we compute the EM power produced mainly by the orbital motion, $L_{\rm EM}^{\rm orbital}$, by decomposing $\Phi_2$ into spherical harmonics and keeping just the $m = 2$ modes for the surface integration. The idea is that $L_{\rm EM}^{\rm orbital}$ captures the energy released in all $l$ modes (we calculate up to $l = 8$) but by excluding the other modes, it avoids the spurious contribution from the background field. Because at merger the luminosity transitions from $m = 2$ to $m = 0$, $L_{\rm EM}^{\rm orbital}$ vanishes at this stage.

Second, we measure the collimated contribution, $L_{\rm EM}^{\rm collimated}$, to the EM power by integrating only where we estimate the jet to be. We therefore integrate over just a solid angle centered on the $z$-axis of width 15 degrees. This luminosity does not vanish after the merger, but rather it tends to the one corresponding to a single spinning black hole as studied in the previous section with spin $a \simeq 0.67$ (which corresponds to the final black hole).

We can now obtain our estimate of the non-collimated EM luminosity by subtracting these two luminosities, i.e. $L_{\rm EM}^{\rm isotropic} \approx L_{\rm EM}^{\rm orbital} - L_{\rm EM}^{\rm collimated}$. Examination of the figure shows that the non-collimated contribution is smaller than the collimated one for most of the early stage of the evolution. At the merger stage however, a strong burst is produced radiating energy in all directions signaling tantalizing prospects for electromagnetic counterparts to gravitational waves, possibly observable regardless of the jet direction.

## V. FINAL COMMENTS

We have studied the interaction among a black hole, the tenuous plasma in its vicinity, and a magnetic field anchored by a circumbinary disk, extending the work of BZ who work under assumptions of stationarity and axisymmetry. Working with dynamical gravity with no assumed symmetries, we find that orbiting, non-spinning binary black holes produce collimated tubes of Poynting flux for each black hole in addition to the expected gravitational wave output. We then remove all angular momentum by considering the head-on collision of two black holes. Remarkably, this system also generates Poynting flux associated with each black hole, suggesting that the motion of the black hole relative to the asymptotic magnetic field direction will generically produce Poynting flux in what might be considered a generalization of the BZ mechanism.

We also study single black holes, both as tests of our code and as simpler physical systems than the binary cases. We study the dependence of the luminosity on black hole spin and alignment. Even when the black hole spin is perpendicular to the asymptotic magnetic field, it still produces a non-zero electromagnetic luminosity.

An important aspect of this work concerns the implications for observing such binary systems. The collimated Poynting flux can be expected to accelerate charges in the vicinity of the black hole which will then radiate copiously through synchrotron processes. There is also the possibility that the radiation will interact with surrounding matter in way that might be observable.

That the EM fields have a clearly discernible pattern tied to the dynamics of the system, makes them possible *tracers of the spacetime* –in the electromagnetic sector– physical characteristics of the systems might be discernible from observable EM signals. In particular, the Poynting flux from the merger resembles the "pair of pants" picture of the event horizon for such a merger [49], with the important difference that the Poynting flux is potentially observable. The collimated flux of energy displays a twisting behavior directly tied to the orbital motion of the individual black holes suggesting the remarkable possibility of "seeing" strong-field gravity in action. Even after the merger when the gravitational waves cease, the Poynting flux continues as per the original BZ mechanism.

Black holes interacting with plasmas may constitute ideal systems for coincident detection in both gravitational and electromagnetic spectra. Deciphering the combination of information obtained in both bands will allow unprecedented scrutiny of strongly gravitating and highly dynamical systems. At a more speculative level, such combined signals might be exploited to shed light on alternative theories of gravity in which photons and gravitons might propagate at different speeds or gravitational energy could propagate out of our 4-dimensional brane (for a recent discussion of some possibilities see [50, 51]).

### Acknowledgments

It is a pleasure to thank J. Aarons, A. Broderick, P. Chang, B. MacNamara, O. Sarbach, A. Spitkovsky and C. Thompson as well as our long time collaborators M. Anderson, E. Hirschmann, P. Motl, M. Megevand and D. Neilsen for useful discussions and comments. We acknowledge support comes NSF grant PHY-0803629 to Louisiana State University and PHY-0803624 to Long Island University as well as NSERC through a Discovery Grant. Research at Perimeter Institute is supported through Industry Canada and by the Province of Ontario through the Ministry of Research & Innovation. Computations were performed Teragrid and Scinet. CP and LL thank the Princeton Center for Theoretical Physics for hospitality where parts of this work were completed.

## VI. APPENDIX

### BZ mechanism. Exploiting symmetries

One common way to interpret the BZ mechanism exploits the time-symmetry of the problem to examine the rate of energy flux from the black hole [42]. Unfortunately, the energy flux is not positive definite and thus the associated fluxes need to be interpreted with care. We briefly review the main details of this approach below.

For any stationary axisymmetric system, one can define a conserved flux vector from the energy conservation equation

$$\nabla_b(\xi_a T^{ab}) = 0 \ . \tag{27}$$

The conserved electromagnetic energy flux is constructed with the time killing vector (i.e., stationary spacetime) $\xi^t = (1, 0, 0, 0)$. The conservation Eq. (27) implies that the radiated energy crossing a spherical surface at a given radius is

$$\partial_t E = 2\pi \int_0^\pi \sqrt{-g} F_{EM} d\theta \quad , \quad F_{EM} \equiv -T^r_t \ . \tag{28}$$

Assuming Kerr-Schild coordinates, we compute the energy flux density $F_{EM}$

$$\begin{aligned} F_{EM} =& \ 2 \left(B^r\right)^2 r \, \Omega_F \left(\frac{a}{2\,M\,r} - \Omega_F\right) \sin^2 \theta \quad (29) \\ & - B^r B^\phi \, \Omega_F \, \Delta \, \sin^2 \theta, \end{aligned}$$

where $\Delta = r^2 + a^2 - 2\,M\,r$. This expression simplifies at the horizon since $r = r_+ = r_H$ and $\Delta = 0$ so that it becomes

$$F_{EM}|_{r=r_H} = 2 \left(B^r\right)^2 r_H \, \Omega_F \left(\Omega_H - \Omega_F\right) \sin^2 \theta, \tag{30}$$

where $r_H = M + \sqrt{M^2 - a^2}$ is the radius of the horizon. This result implies that if $0 < \Omega_F < \Omega_H$ and $B^r \neq 0$, then there is an outward directed energy flux at the horizon; rotational energy is being extracted from the black hole due to the magnetic field lines. The use of Kerr-Schild coordinates allow for direct computations of the flux at the horizon without any special treatment. However, as mentioned, one message from this calculation is that energy comes out of the event horizon which is forbidden at the classical level. The problem lies in the fact that the energy defined with the killing vector $\xi$ is not positive definite within the ergosphere. Consequently this effect is interpreted as negative "killing" energy falling into the horizon.

### BZ mechanism. Exploiting an analogy

A second standard treatment of the problem relies on the membrane paradigm which treats the horizon of the black hole as a fictitious membrane. A thorough discussion of this treatment is presented in [34] and we refer the reader to it for details. Here we briefly mention its key features. This membrane is regarded as having surface charge density, current and resistivity. In particular, the surface resistivity is equal to the vacuum impedance ($R_H = 377\Omega$). Within this framework, it is natural to imagine a circuit with wires parallel to magnetic field lines in the vicinity of the black hole and connected to some load at far distances. To examine the power released by such a circuit in which the rotating horizon of the black hole plays the role of a battery. one needs to find the electromotive force of the black hole. Employing Faraday's law of induction in such circuit one obtains

$$\Delta V \simeq (2\pi)^{-1} \Omega_H B^r r_H^2 \ , \tag{31}$$

which can be used to estimate the total current flowing through the circuit as

$$I \simeq \frac{\Delta V}{\Delta R_H + \Delta R_L}, \tag{32}$$

where $\Delta R_H \simeq R_H$ and $\Delta R_L \simeq \Omega_F/I$ are the resistances along the short end of the circuit near the horizon and at far distances (the load), respectively. With these, the total power dissipated by the circuit can be estimated as

$$P \propto \Omega_F \left(\Omega_H - \Omega_F\right)(B^r)^2 \ . \qquad (33)$$

Notice that this analogy can be extended to argue an electromotive force will be induced also if a black hole moves through an (asymptotically) stationary magnetic field configuration. Thus explaining the Poynting fluxes we see in both the head-on and orbiting binary black hole cases.

Last, a recent work [52] presents an alternative point of view which relies on regarding the region surrounding the black hole as an electromagnetically active medium and re-interprets the extraction process in a way closely tied to Penrose's extraction process from a rotating black hole.

These different approaches to the BZ mechanism agree in their main message. Namely that a net flux of electromagnetic energy is produced by the system the magnitude of which scales as $\propto (B^r)^2 r_H \Omega_F (\Omega_H - \Omega_F)$. This flux is powered by the rotational energy of the spinning black hole.